\documentclass[fleqn,10pt]{wlscirep}
\usepackage[utf8]{inputenc}
\usepackage[T1]{fontenc}
\usepackage{bbold}
\usepackage{graphicx}
\graphicspath{ {Figures/} }
\usepackage{amssymb}
\usepackage{amsmath}
\usepackage{bm}

\usepackage{empheq}
\usepackage{amsthm}
\usepackage{siunitx}
\usepackage{float}
\usepackage{color}
\usepackage{cite}
\usepackage{url}
\usepackage{caption}
\usepackage{dcolumn}

\title{Octet lattice-based plate for elastic wave control}

\author[1,*]{Giulia Aguzzi}  
\author[1]{Constantinos Kanellopoulos}
\author[2]{Richard Wiltshaw}
\author[2,3,4]{Richard V. Craster}
\author[1]{Eleni N. Chatzi}
\author[1]{Andrea Colombi} 

\affil[1]{Department of Civil, Environmental and Geomatic Engineering, ETH Z\"urich, Z\"urich 8093, Switzerland}
\affil[2]{Department of Mathematics, Imperial College London, London SW7 2AZ, United Kingdom}
\affil[3]{Department of Mechanical Engineering, Imperial College London, London SW7 2AZ, United Kingdom}
\affil[4]{UMI 2004 Abraham de Moivre-CNRS, Imperial College London, London SW7 2AZ, United Kingdom}

\affil[*]{aguzzi@ibk.baug.ethz.ch}

\begin{abstract}
Motivated by the importance of lattice structures in multiple fields, we investigate the propagation of flexural waves in a thin woven plate augmented with two classes of metastructures for wave mitigation and guiding, namely metabarriers and metalenses. The cellular architecture of this plate invokes the well-known octet topology, while the metadevices rely on novel customized octets either comprising spherical masses added to the midpoint of their struts or variable node thickness. 

We numerically determine the dispersion curves of a doubly-periodic array of octets, which produce a broad bandgap whose underlying physics is elucidated and leveraged as a design paradigm, allowing the construction of a metabarrier effective for inhibiting the transmission of waves. More sophisticated effects emerge upon parametric analyses of the added masses and node thickness, leading to graded designs that spatially filter waves through an enlarged bandgap via rainbow trapping. Additionally, Luneburg and Maxwell metalenses are realized using the spatial modulation of the tuning parameters and numerically tested. Wavefronts impinging on these structures are progressively curved within the inhomogeneous media and steered toward a focal point. Our results yield new perspectives for the use of octet-like lattices, paving the way for promising applications in vibration isolation and energy focusing.
\end{abstract}
\begin{document}

\flushbottom
\maketitle

\thispagestyle{empty}

\section*{Introduction}
\label{sec1}
\textit{Lattice materials} are engineered periodic structures whose cellular architecture originates from a network of entangled structural elements (e.g., rods, beams, or plates) \cite{Gibson1997}. Their high strength-to-weight ratio, combined with recent progress in additive manufacturing, renders them particularly attractive for the scientific community; their unique mechanical performance has been extensively explored \cite{Maconachie2019}, with broad application in biomedicine, aerospace and energy absorption. It is only recently that these have drawn attention for their ability to diffract waves and behave as filters \cite{Phani2006} that forbid energy transmission across selected frequencies, known as \textit{bandgaps} \cite{Brillouin1953}. The periodic nature of these reticulated structures is often at the root of their attenuation mechanism, leading to Bragg scattering effects \cite{Simon2013} that dominate the spectrum of their band structure \cite{Delpero2016}. Bragg bandgaps arise through destructive interference, occurring when the wavelength of a wave is comparable to the fundamental period of the media they are propagating through, and typically occur at frequencies too high to be exploited. Much work has been done to broaden Bragg bandgaps by introducing locally resonant objects, whose resonance coincides with the Bragg frequency \cite{kaina2013composite, taubert2012octave, goldberg2009exciton}, wherein lattice vectors are carefully chosen such that these frequencies coincide. 

Despite offering wave control, Bragg-based frame designs suffer from a periodicity requirement that forces their characteristic size to comply with the targeted wavelength \cite{Brillouin1953} thus, limiting their role. To overcome this limitation, attempts are being made to design lattices with resonance-induced bandgaps that, unlike their phononic counterpart, exploit the coupling between propagating waves and resonating components and offer greater tuning flexibility at subwavelength scale (i.e., the size of the unit cell can be orders of magnitude smaller than the wavelength). These structures form a subset of \textit{single-phase metamaterials} \cite{Gonella2009, Liebold2014,Junyi2016} that use auxiliary resonators or inborn resonant modes of the lattice to hybridize propagating waves and inhibit their transmission. Common practice for their implementation includes, for instance, embedding stand-alone resonant elements (e.g., cantilever beams \cite{Gonella2009}) into an original frame-based topology, similarly to the solution proposed by Li \textit{et al}. \cite{Li2017} 
A different method, adopted by An \textit{et al}. in \cite{An2020} uses, instead, the constitutive matrix of a body-centered cubic (BCC) lattice and constructs an integrated resonator via radius jump of its strut cross-section. The BCC cell is only one of the Bravais lattices \cite{Ashcroft1976, Simon2013} potentially serving as metamaterials. Among the fourteen types known to exist in the three-dimensional space \cite{Simon2013}, the \textit{octet} design \cite{Deshpande2001} is an example of face-centered cubic (FCC) topology with promising attenuation potential. The dispersive behavior of this lattice, already renowned in statics for its high strength combined with a slender lightweight profile \cite{Ashby2005}, has been outlined by Arya \textit{et al}. \cite{Arya2011} who show that the octet supports a bandgap whose existence depends upon the aspect ratio of the struts, but do not investigate the underlying physics of its generating mechanism; only briefly suggested by Gerard \textit{et al.} \cite{Gerard2021}. In addition, most of the state-of-the-art research on such a FCC cell resort to frequency gaps produced by attached resonators \cite{Junyi2016} or use it as a constitutive matrix of multi-phased materials  \cite{Chen2014, Arretche2018}, without delving into its inherent resonant modes. The absence of an overarching dynamic assessment of this architecture in the existing literature emphasizes the novelty of our work, which proposes the octet topology as a metamaterial lattice endowed with an energy gap stemming from the bending local resonance of its beam-like members. Moreover, due to the arrangement and behavior of wave propagation through the struts, we show that the octet cells form a medium in which flexural resonances naturally coincide with the Bragg frequency - as observed within the eigenmodes permitted by Floquet-Bloch waves.

In aerospace and aircraft engineering, lattice plates are frequently used as structural components or filling in sandwich panels \cite{Gibson1997} that, whilst providing lightness to the structure \cite{Maconachie2019}, are liable to significant vibrations. The propagation of such vibrations in solid thin plates, occurring in the form of Lamb waves with different polarization, has been extensively documented \cite{Lamb1917,Graff2012} resulting in a variety of metamaterial-based solutions. Prominent among these are \textit{metabarriers} (or metawedges \cite{Colombi2016}) encompassing clusters of rods whose longitudinal resonant mode couples with the flexural waves propagating in the plate and triggers the opening of bandgaps \cite{Rupin2014,Colombi2017elastic}. To enlarge the frequency band of such barriers, that owing to local resonance offer a fairly limited range of applicability, \textit{graded} designs have been explored, where the spatial variation of the tuning parameter, rod height, yields a wider spectrum of attenuation associated to the so-called Rainbow trapping effect \cite{Tsakmakidis2007,Deponti2020}. The theory underpinning these metastructures can be easily applied to octet-based plates, with the longitudinal resonances replaced by the bending mode of the struts and the rod height by the parameters of the beam-like members. Based on these considerations, a number of parameters exist in the literature to tune the resonance-bandgap, two of which are of particular interest for the present study, namely auxiliary point masses and joint stiffness. Junyi \textit{et al}. \cite{Junyi2016} proposed equipping the end of cantilever beams, added to primary $3$D Bravais lattices, with point masses whose size governs the gap frequency independently of the geometry of the base lattice. Shortly before this, the same spherical masses were attached to the joints of a Kagome lattice by Liu \textit{et al}. \cite{Liu2014} to stiffen its connections and open a bandgap tied to the rod-node vibration.

Beyond widening the bandgaps, graded designs lay the foundation for a multitude of compelling wave control possibilities. One of these dates back to the nineteenth century and the seminal work of Maxwell on so-called flat or \textit{gradient-index} (\textit{GRIN}) \textit{lenses} \cite{Maxwell1854}. These are effective, aberration-free devices that leverage composite structures to adjust the spatial distribution of the refractive index within a bounded region and ultimately shape the incident wavefronts. The \textit{Luneburg lens} \cite{Luneburg1944} is a type of spherical GRIN lens that diverts any collimated beam colliding at its boundaries to a diametrically opposite focal point. Similarly to this system, underpinned by a refractive index that increases radially from the outer toward the center of the graded area, the \textit{Maxwell fish-eye lens}, bends the trajectories of rays originated by a point source at its surface and focuses them in a mirrored image point. While building such lenses with ordinary composite materials appears a daunting task, employing phononic crystals \cite{Climente2014,Tol2017} or metamaterials \cite{Lee2018,Park2018} considerably relieves this complexity. From a metamaterial perspective, the cluster of locally resonant rods with graded height, previously discussed, has proven an effective candidate for the design of GRIN lenses \cite{Colombi2016lens, Zhu2017,Dominiguez2021}. Its rod-resonance bandgap, besides providing attenuation, serves as tuning variable for the phase velocity of waves traveling in the metastructure and leads to applications at different scales, from acoustics \cite{Colombi2016lens} to seismology \cite{Colombi2016transformation}. In terms of frame structures, the research on wave guiding is still at a primordial state, mainly confined to two-dimensional lattices \cite{Ruzzene2003,Casadei2013,Pal2016,Zelhofer2017} and yielding, only in a handful of cases, to more sophisticated effects. Xie \textit{et al}. \cite{Xie2018} and Zhao \textit{et al}. \cite{Zhao2020} were among the few to design $2$D and $3$D flattened Luneburg lenses consisting of a three-dimensional truss-based square lattice, to manipulate waves in the ultrasound regime. A recent study \cite{Bayat2018}, on the other hand, has explored the propagation of elastic waves in three-dimensional lattices, including specifically the octet cell. Although similar to our investigation with respect to the choice of topology, no bandgap is reported in this case and only the anisotropic character of the structure is emphasized, with no explicit reference to shaping wave trajectories via graded octet lattices. Therefore, the lack of existing literature on gradient-index lenses designed with octet lattices further motivates our work.

Here we take the octet lattice analysis one step further by using this geometry to realize a reticulated plate and two types of integrated devices for flexural wave mitigation and control. 
Following a dispersion analysis of the proposed standard octet cell, the bandgap and its underlying generation mechanism are revealed, resulting in the identification of two tailoring parameters. Auxiliary masses, embedded in the midpoint of the struts, or variable joint thickness are used to tune the frequency range with forbidden propagation as well as the phase velocity spatially within the lattice. 
To avoid redundancy of results, only those stemming from the addition of masses to the primary cell are reported here; varying the node thickness produces similar outcomes. Two metabarrier designs are discussed, the first characterized by constant auxiliary masses and the second relying on the grading strategy to trap and hinder wave transmission over a wider frequency band. Finally, Luneburg and Maxwell lenses are built by smoothly increasing the supplementary spherical masses radially from the boundaries to the center of the devices. We find the octet cell and subsequent customizations possess desirable dispersive properties easily coaxed to required frequencies, these properties are readily exploited to create a variety of devices able to manipulate the propagation of waves over a variety of scales, in both a subwavelength (lenses) and Bragg (metabarriers) scattering regime.  

\begin{figure*} [t]
    \hspace*{-0.5cm} 
	\centering
		\includegraphics[scale=1]{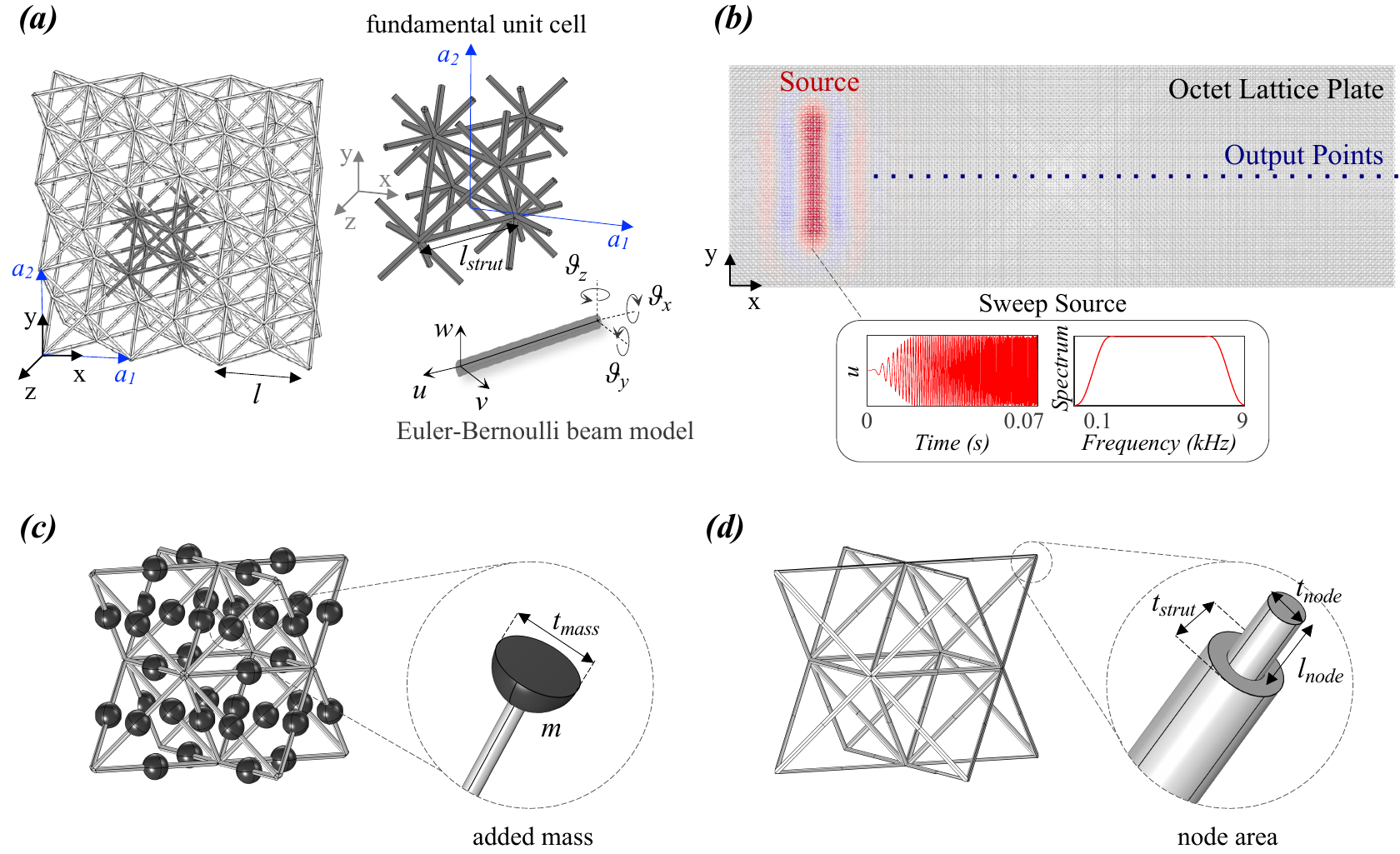}
	\caption{Cell designs and setups for dispersion and parametric analyses.
	(a) $2$D standard octet lattice with fundamental unit cell depicted in the zoom ($a_{1}$ = ($l$, 0, 0) and $a_{2}$ = (0, $l$, 0)).
	(b) Finite lattice plate ($3.6$ x $1.2$ x $0.03$ m) for the validation of the band structure via time-transient simulations in Real-ESSI. The constitutive unit corresponds to the standard octet reported in (a). The plane sweep source and line of points where the output signal is recorded, spaced $0.03$ m apart and located on the nodes of the lattice, are marked in red and blue. (c) Customized octet equipped with spherical masses $m$, of diameter $t_{mass}$, (dark color) added to the midpoint of the struts. (d) Octet with customized joints via tunable node thickness, $t_{node}$. The zoomed plot inset shows the node area of length $l_{node}$ equal to $2t_{strut}$ and of variable thickness $t_{node}$. }
	\label{FIG:SystemDR}
\end{figure*}

\begin{figure*}[t]
    \hspace*{-1.1cm} 
	\centering
		\includegraphics[scale=1]{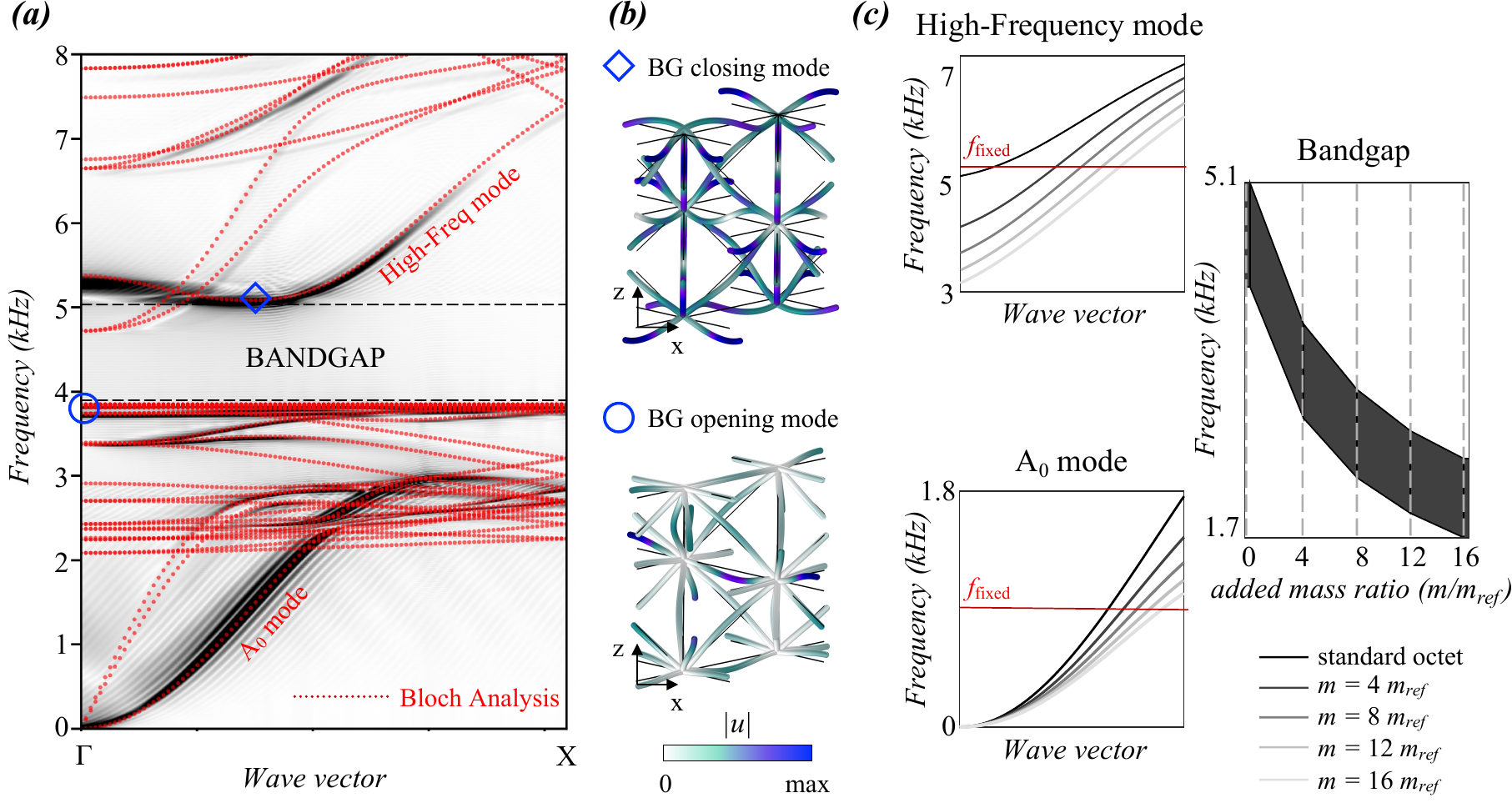}
	\caption{Dispersion analysis. (a) Band structures of the standard octet crystal and finite plate (Figures \ref{FIG:SystemDR}a, b) computed with Comsol Multiphysics (red) and Real-ESSI (black). The $\Gamma$-$X$ wavevector in the reciprocal space represents waves propagating in the  $\textbf{e}_x$ direction in the physical system. The red curves are obtained via Bloch-eigenvalue analysis on the fundamental unit cell of Figure \ref{FIG:SystemDR}a. The black branches derive from the fast Fourier transform (FFT) of the $z$-wise signal recorded on the receivers of Figure \ref{FIG:SystemDR}b, where they are denoted as output points and colored in blue.  
	(b) Bandgap (BG) opening and closing eigenmodes, respectively at $3.85$ and $5.08$ kHz. (c) Sensitivity analysis on the dispersion curves of the customized octet cell (Figure \ref{FIG:SystemDR}c) for increasing $m$, where $m_{ref} $ = $ \rho A 2t_{strut} $.}
	\label{FIG:DR}
\end{figure*}

\section*{Results}
\subsection*{Octet cell design, dispersion and parametric analyses}
\label{sec2}
Consider a thin reticulated plate, of infinite extension, constructed from repeating unit cells, each comprising a strut network generating the well-known octet topology \cite{Deshpande2001}. We choose a fundamental cell whose edges coincide with the midpoint of the octet struts, as depicted in Figure \ref{FIG:SystemDR}a, where we define physical basis vectors $\textbf{a}_1$ and $\textbf{a}_2$, assumed collinear with the Cartesian axes $x$ and $y$. 
We consider struts of circular cross-section, of diameter $t_{strut} = 1$ mm, Young's modulus $E$ = $1.7$ GPa, density $\rho$ = $450$ kg/$m^3$ and Poisson's ratio $\nu$ = $0.3$. We take the length of our lattice vectors to be $l = 3$ cm and hence have a strut length of $l_{strut} = l \sqrt{2} / 2$. 
The circular cross-section provides axial symmetry and prevents stress concentration that would instead arise in sharp-edged profiles, e.g., square cross-section.   

It is sufficient to consider waves propagating through the fundamental cell as the phase shift from cell-to-cell is completely described by the dispersion relation between eigenfrequencies and Bloch-wavevector, corresponding to the red dotted line reported in Figure \ref{FIG:DR}a (see Methods). 
Despite the complexity of this folded band structure, three main frequency bands can be clearly identified.

In the low-frequency limit, from $0$ to $2$ kHz, the spectrum is dominated by zero-order modes similar to the solutions of Lamb's characteristic equations in solid, homogeneous, and thin plates for long wavelengths \cite{Lamb1917}. The underlying lattice structure has subwavelength behavior and wave propagation closely resembles an equivalent continuum supporting three propagating waves, respectively two with transverse and one with longitudinal polarization \cite{Pennec2008}. Among these, the asymmetric flexural mode, simply denoted as \textit{$A_{0}$ mode} in Figure \ref{FIG:DR}a, is the most attractive for its ease of activation and amount of carried energy hence, the cornerstone of this study.

At higher frequencies (2 to 5 kHz), the wavelength reduces rapidly and waves interacting with the detailed geometry of the cell open a \textit{bandgap} between $3.85$ and $5.08$ kHz. Waves with frequency content matching this range are prevented from penetrating the octet lattice while being trapped by the bending vibrations of its struts. These deformations emerge as flat bands in the lower bound of the forbidden interval and serve as evidence of its local resonance-based generating mechanism \cite{Goffaux2002}. They cluster around the first flexural frequency of a clamped-clamped Euler-Bernoulli (EB) beam, geometrically equivalent to the struts in the octet and defined by:
\begin{equation}
    f_{clamped} = \frac{22.4}{2\pi l_{strut}^2} \sqrt{\frac{\mathrm{EI}}{\rho \mathrm{A}}}
    \label{eq:FixedStrutResFreq}
\end{equation} 
 The analytical flexural frequency in Equation (\ref{eq:FixedStrutResFreq}), equal to $3.85$ kHz, corresponds to the bandgap opening bound and poses as a design variable for tailoring the bandgap frequency range. By considering the octet struts as equivalent EB beams with clamped edges and bending resonance \cite{Liebold2014}, the bandgap opening frequency can be estimated via Equation (\ref{eq:FixedStrutResFreq}) and, consequently, also the correlated attenuation zone \cite{Wang2015}. This simplification reduces the complexity of the octet cell to individual beams clamped at the nodes, facilitating  the tuning of design parameters. This clamped-clamped nodal behavior relies on a network of highly connected struts peculiar of the octet topology. As outlined in Wang \textit{et al}. \cite{Wang2015} highly connected lattices (here intersection points of 12 members at a node) yield a clamped-clamped behavior of the struts at the nodes. 
 Similarly to Nolde \textit{et al}. \cite{nolde2011high}, our flat bands at $f_{clamped}$ correspond to standing waves induced by the fundamental resonance of vibrations along clamped-clamped beams. This  behavior is highlighted in the BG opening eigenmode displayed in Figure \ref{FIG:DR}b, where the nodes stand still. Furthermore, such a fundamental resonance fits exactly within the length of the primitive lattice vectors (a half our conventional unit cell), hence the propagation regime is identical to Kaina \textit{et al}. \cite{kaina2013composite} in which the wavelength of the fundamental clamped-clamped resonance is twice that of the modulus of the primitive lattice vectors; subsequently the Bragg frequency and $f_{clamped}$ coalesce, resulting in destructive interferences producing a broad bandgap. Wang \textit{et al}. \cite{Wang2015} remarks that bandgaps, in this setting, are generated by local resonances and not Bragg scattering; however, in the case where beams are parallel to and of the same length as the primitive lattice vectors, one expects a medium in which the fundamental resonances naturally match the Bragg frequencies, all of which are properties the underlying octet cell possesses.
 
Beyond the bandgap, approximately from $5$ to $8$ kHz, additional dispersive modes populate the spectrum of the dispersion relation, mimicking waves that propagate at the strut level accompanied by roto-translations of the nodes. In particular, a branch with predominant orientation along $z$, denoted as \textit{high-frequency mode} in Figure \ref{FIG:DR}a, can be observed. Owing to its polarization, this defines the upper bound of the attenuation zone (see top inset in Figure \ref{FIG:DR}b), in contrast to the two modes converging at $4.7$ kHz, which exhibit, instead, in-plane deformations. 

One known limitation of the Bloch analysis is that only a minority of the computed modes are triggered owing to their strong source dependence. Hence, a finite octet-based plate, consisting of $120$ x $40$ cells equivalent to the one of the infinite study (Figure \ref{FIG:SystemDR}b), is adopted to corroborate the predicted band structure via time-domain simulations (see Methods). 
The outcome of this analysis is shown as a background of the dispersion curves in Figure \ref{FIG:DR}a, where the branches involving out-of-plane deformations are predominantly excited. Almost complete superposition can be appreciated between the $A_0$ and high-frequency modes of the frequency-wavenumber spectrum and that of the Bloch analysis. A small discrepancy arises only in the higher-order modes and in the upper bound of the bandgap, with a downward shift of about $0.05$ kHz in the results of the time-transient simulation. From 2 to 4 kHz, additional modes, generated by waves with polarization in the lattice plane (longitudinal or shear), are effectively captured by the finite model. Overall, the results of these two types of analyses reveal good agreement and justify their joint application in the following sections. 

\begin{figure*}[t]
	\centering
 	\hspace*{-0.5cm} 
		\includegraphics[scale=1]{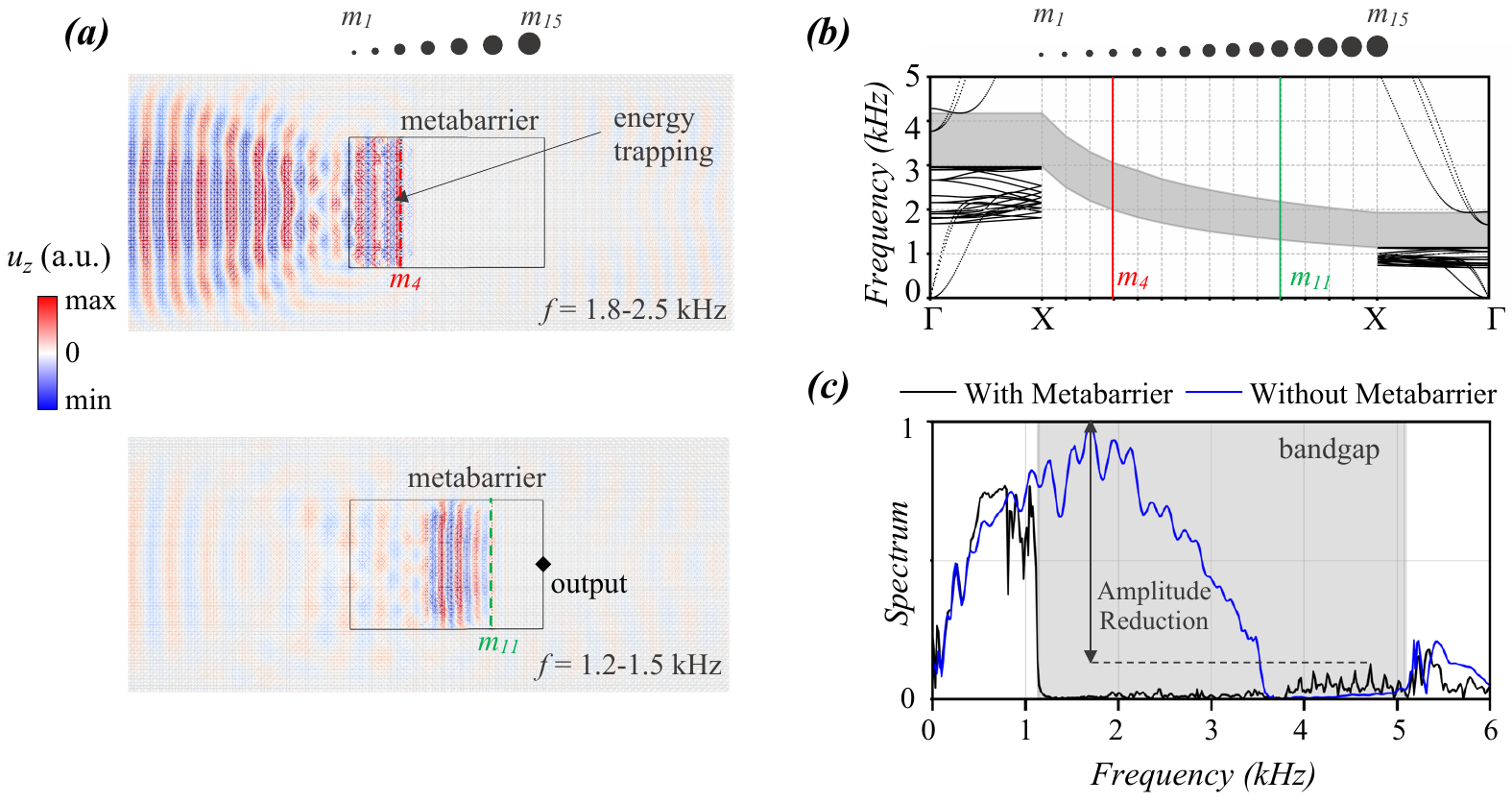}
	\caption{Temporal and spectral wave fields of the graded metabarrier. (a) The top and bottom insets show the displacement field filtered within the bandgaps induced by octets with masses $m_{4}$ = $11m_{ref}$ (red) and $m_{11}$ = $31m_{ref}$ (green). (b) Evolution of the bandgap within the graded metabarrier, for masses increasing from $m_{1}$ to $m_{15}$, computed via Bloch analysis (see Methods). (c) The spectrum of a signal recorded on the right end of the metabarrier (black), in the output point emphasized in (a), is compared with that of a receiver located at the same position in a bare lattice plate (blue). The bandgap generated by this graded design is highlighted in gray. The amplitude reduction within the bandgap is estimated as the difference between the maximum spectrum of the two signals in the attenuation zone, as illustrated in the figure. }
	\label{FIG:MetabarriersGraded}
\end{figure*}

The octet unit outlined hitherto, comprising 24 woven struts of constant, circular cross-section, is denoted as the \textit{standard cell}. Remarkably, a local resonance-induced bandgap originates in this single-block lattice owing to the refined topology of its design, thus paving the way for a variety of parameters to be leveraged in tuning the dispersion relations. While geometrical and material properties emerge as the primary natural candidate, their influence on the octet band structure has already been assessed \cite{Arya2011},\cite{Aguzzi2020}. The focus here shifts to two novel, so-called, \textit{customized octet cells} whose parameters allow tailoring the band structure without affecting the original geometry, i.e., cell width and strut thickness. In the customized cell of Figure \ref{FIG:SystemDR}c the midpoint of the struts is endowed with auxiliary spherical masses, $m$, of diameter $t_{mass}$, as this undergoes major displacements when bending deformations are activated. In order to retain a monolithic structure, easier to 3D print with modern techniques, the supplementary mass is made of the same material selected for the lattice and is expressed as a multiple of a reference value, $m_{ref} = \rho A 2 t_{strut} $. The diameter $t_{mass}$ in Figure \ref{FIG:SystemDR}c is then derived upon definition of this quantity and its practical feasibility, with respect to the thickness of the struts, is adequately verified in each of the proposed devices. The spheres are added as point masses in the finite element model in Comsol $5.6$ and the sensitivity analysis is performed on the single customized unit via Bloch theory. Likewise the approach adopted for the standard octet cell, the struts are treated as equivalent fixed-fixed Euler-Bernoulli beams with the distributed mass, $\rho$$A$$l_{strut}$, incremented by the point mass $m$. Naturally one expects these added masses to provide a damping over the whole band structure reducing the frequencies of comparable eigenmodes, as observed in Figure \ref{FIG:MetabarriersGraded}b. The first fundamental frequency, governing the position of the bandgap, has an inverse reliance on the added mass, as already analytically proven by Low \textit{et al}. \cite{Low1997}. When $m$ increases, the bandgap is instantly shifted to lower frequencies, from the range $3.85$-$5.08$ kHz to $1.73$-$2.74$ kHz in the rightmost inset of Figure \ref{FIG:DR}c, altering the $A_0$ and high-frequency modes. Each point in these dispersion curves has an associated phase velocity, $v_p = \omega/ k_{x}$ \cite{Simon2013}, which decreases when moving from lighter to heavier masses at constant frequency (red line in Figure \ref{FIG:DR}c), thereby animating the lattice with slower waves.  
Similar outcomes arise when exploring the customized cell in Figure \ref{FIG:SystemDR}d. Following the analogy between octet struts and clamped EB beams, we harness the topology of the connections in the lattice to modify the boundary conditions of the equivalent beams and their natural resonant frequency, in turn affecting the band structure of the architected cell. A node area of length $l_{node}$, twice as long as $t_{strut}$, is identified in Figure \ref{FIG:SystemDR}d, where the diameter of the section, $t_{node}$, is expressed as a fraction of the original thickness $t_{strut}$. By decreasing this parameter from $t_{strut}$ to $0.2t_{strut}$, clamped-like joints are replaced by quasi-hinged connections that allow almost free rotations between contiguous struts. This yields a substantial reduction of the node bending stiffness, which is proportional to the moment of inertia, $I$, and consequently to the joint thickness, $t_{strut}^4$. The bandgap opening frequency is therefore shifted from the value reported in Equation (\ref{eq:FixedStrutResFreq}) to that of a simply supported beam, $f_{pinned} = 0.44f_{clamped}$, entailing a slope-related change in the low ($A_0$) and high-frequency branches. Again, the phase velocity of the waves diminishes, leading to slower waves propagating in the hinge-like octet lattice and faster waves being instead transmitted by the stiffer, clamped-based frames. 

In designing the cell of Figure \ref{FIG:SystemDR}d, compliant joints were preferred over decoupling the struts with prefabricated hinge-connectors as they preserve the single-phase material while allowing a gradually varying stiffness of the connections. Although beyond the scope of this study, it is important to point out that this choice may cause a high stress concentration in the joints that must be examined prior to fabrication. 

\begin{figure*}[t]
	\centering
	\hspace*{-0.8cm} 
		\includegraphics[scale=1]{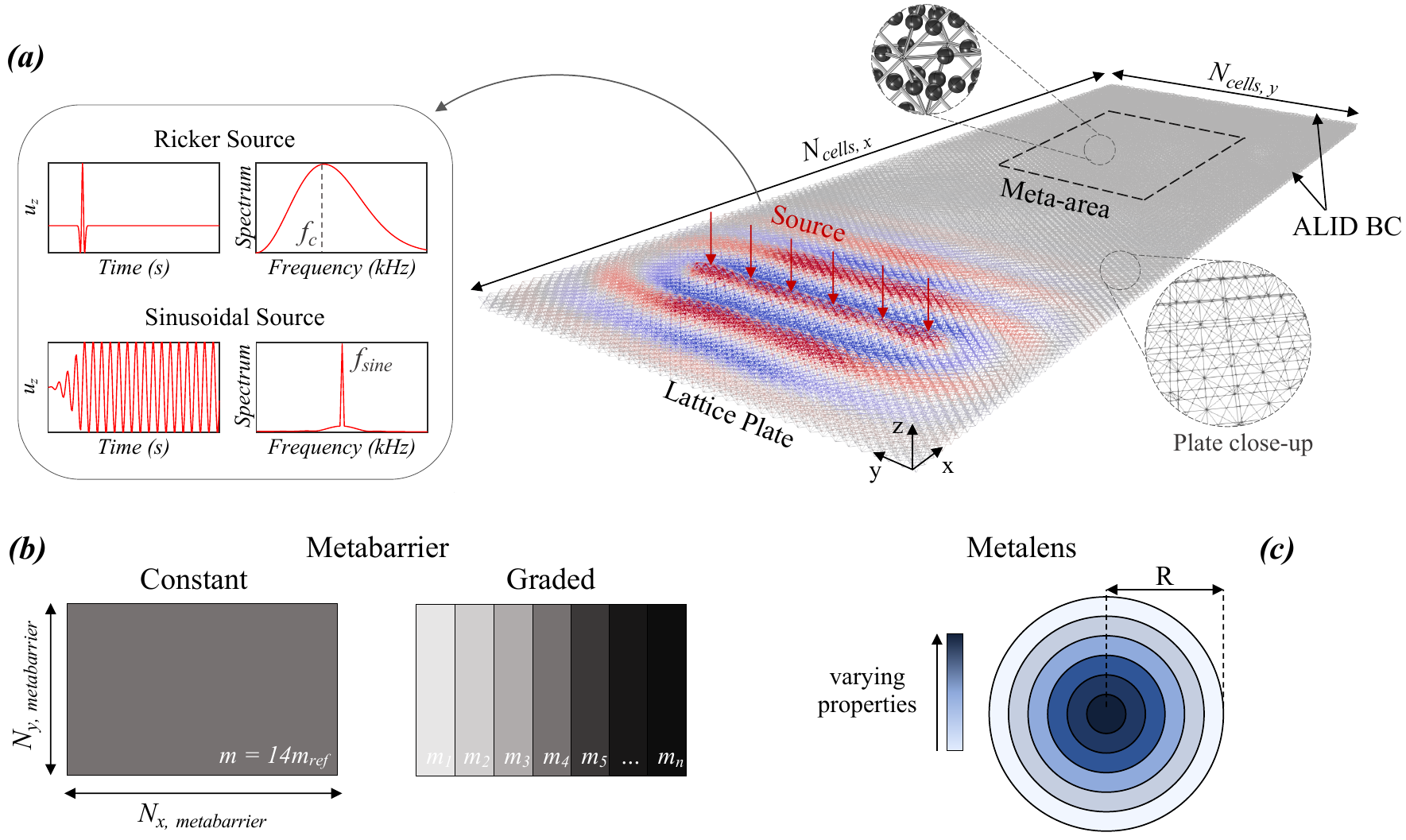}
	\caption{Setup for time-transient simulations in Real-ESSI. (a) Lattice plate made of standard octet units as detailed by the close-up. The meta-area (dashed black line) consists of embedded-mass octets arranged as rectangular metabarriers or circular metalenses, as reported in (b) and (c). Absorbing boundary conditions with quadratically increasing damping \cite{Rajagopal2012} (ALID) are implemented on all four edges of the plate. The two sources used for the metabarriers (Ricker) and metalenses (sine) are sketched in the inset.
	(b) Metabarries of customized cells with auxiliary masses. On the left side, the constant setup ($50$x$20$) encloses cells with $m$=$14m_{ref}$. The graded configuration is formed by $30$ x $20$ cells of progressively increasing masses, $m_n$ with $n=1,...,15$. Each layer comprises two rows of octet cells with identical supplementary mass. (c) Layout for Luneburg and Maxwell metalenses featuring concentric circular layers with variable point masses. In both cases, the lens radius, $R$, consists of $12$ cells of width $l$ and the source is a sinusoidal excitation tuned at the design frequency of the lens.}
	\label{FIG:System}
\end{figure*}

\subsection*{Wave propagation in finite meta-lattices}
\label{sec3}
The computational domain, depicted in Figure \ref{FIG:System}a, is a $3$ cm thick reticulated plate assembled by spatially tessellating the standard octet cell illustrated in Figure \ref{FIG:SystemDR}a. A so-called \textit{meta-area} lies within this structure, where the parameters of the octet, introduced in the previous Section and expressed in terms of supplementary masses or joint stiffness, are progressively tuned to yield exotic attenuation or wave guiding phenomena. Therefore, two types of devices are built, as reported in Figures \ref{FIG:System}b and c: the \textit{metabarriers} and \textit{metalenses}. The metabarriers leverage an arrangement of customized octet units, either tailored at the same or diverse frequencies, to generate an inhibited frequency range, where the displacement field vanishes and ultimately shield a target area. Conversely, the metalens relies upon the spatial variation of lattice properties to modify the refractive index in an inhomogeneous bounded region and manipulate wave trajectories, eventually offering the opportunity to focus and harvest vibration energy. 

\begin{figure*}[t]
\hspace*{-0.8cm}   
		\includegraphics[scale=1]{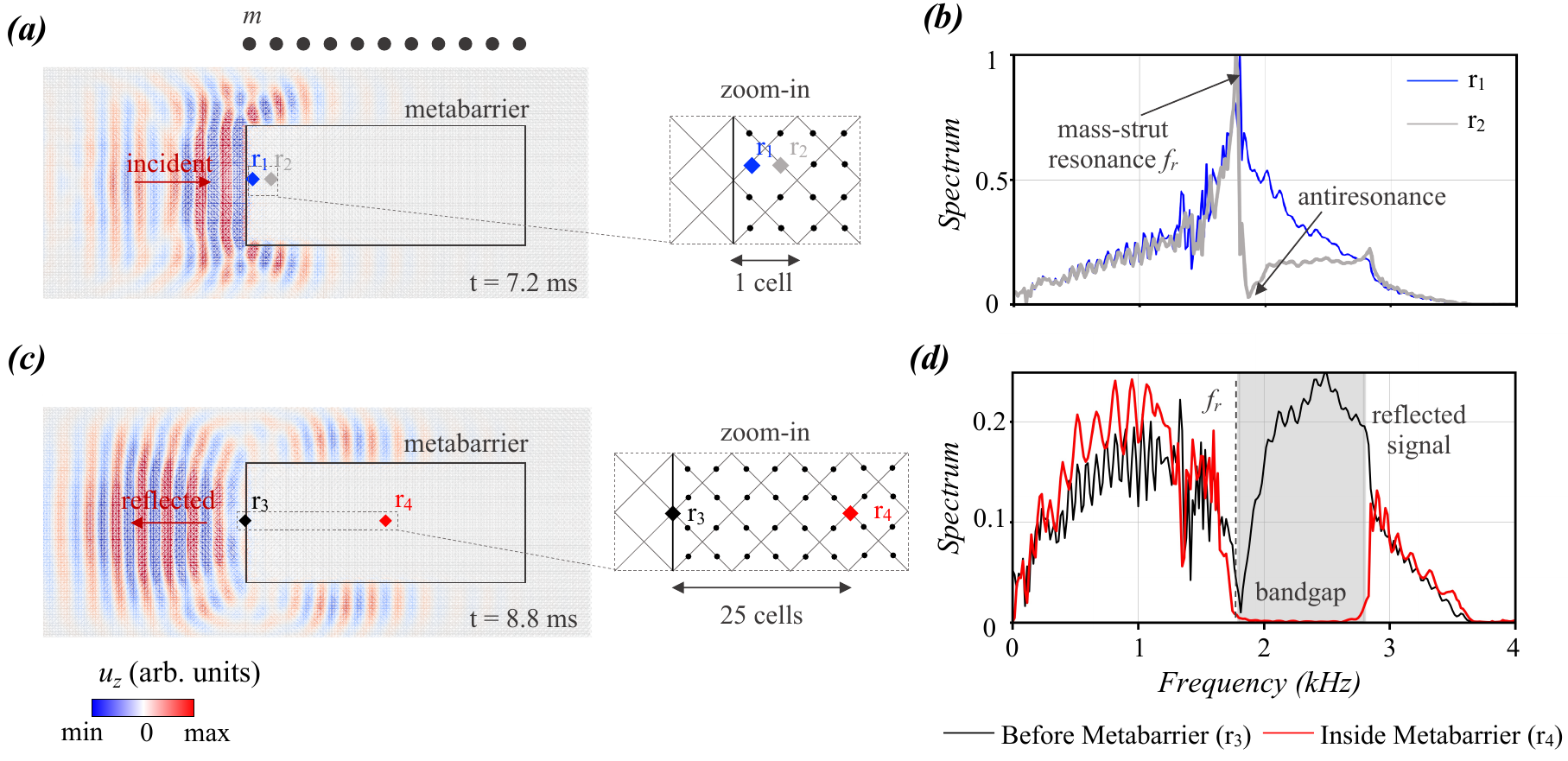}
	\caption{Temporal and spectral wave fields of the constant metabarrier. (a) Propagating Lamb waves filtered within the bandgap ($1.8$ - $2.8$ kHz). The output signal is recorded in two points, $r_1$ and $r_2$, arranged on the midpoint of the struts atop the embedded masses in order to validate the hypothesis of hybridization bandgap. Their spectrum is reported in (b). (c) Filtered displacement field at a later time instant. The back propagating, reflected wavefront is emphasized by the red arrow. Two receivers, located respectively before and within the barrier, are shown in the zoom. Their spectra are compared in (d), after normalization with the peak value of (b).}
	\label{FIG:MetabarrierConstant}
\end{figure*}

\subsubsection*{Wave propagation in lattice metabarriers}
\label{sec31}
Two lattice metabarriers, comprising mass-embedded octets, are implemented within a lattice plate made of $120$ by $40$ standard cells, arranged as per the configuration in Figure \ref{FIG:System}a, and separately investigated in this section.

Illustrated in Figure \ref{FIG:System}b is the $50$x$20$ cells \textit{constant metabarrier}, where the struts of the octets are equipped with midpoint masses, $m$, equal to $14 m_{ref}$ ($t_{mass}$ = $3.5$ mm), yielding an attenuation zone from $1.8$ to $2.8$ kHz (gray region in Figure \ref{FIG:MetabarrierConstant}d). The metastructure is assessed under excitation of a $60$ms Ricker wavelet located at a distance of $45$ cells to guarantee a plane wavefront. The spectrum of the source, with dominant component along $z$, is centered on the bandgap, at $2.1$ kHz.

The recorded displacement field $u_z$, filtered inside the bandgap at two sequential instants in Figures \ref{FIG:MetabarrierConstant}a and c (see video in the Supplementary Material), unveils the attenuation potential of the lattice metabarrier, based on Fano-like resonances \cite{Miroshnichenko2010}. At \textit{t}=$7.2$ ms, the flexural waves scattered by the resonating mass-struts in the metawedge interfere destructively with the unscattered incoming field, impeding the transmission of energy \cite{Lemoult2013}. The resulting energy gap, originated by the hybridization between the asymmetric propagating Lamb wave and the coalescence of the resonant bending mode of the struts and Bragg scattering, induces splitting of the incident wavefront \cite{Zaccherini2020}, as depicted in Figures \ref{FIG:MetabarrierConstant}a, c.

This mitigation mechanism, already introduced in the previous Section, is further corroborated by the spectrum of two waveforms, $r_1$ and $r_2$ in Figures \ref{FIG:MetabarrierConstant}a, b, located on the masses of a cell belonging to the first row of the metabarrier. The signal of the leftmost mass-strut (blue line in Figure \ref{FIG:MetabarrierConstant}b) exhibits a pronounced Lorentzian symmetric peak centered at the resonant frequency $f_{r}$ ($1.8$ kHz), whilst the second waveform, $r_2$, reveals the typical asymmetric profile, where a sharp enhancement of the spectrum prior to $f_{r}$ is followed by a transmission drop (antiresonance) \cite{Miroshnichenko2010}.
This Fano-type interference, limited to a narrow bandwidth if only the first line of cells in the metabarrier is considered, intensifies when leveraging an array of octets tuned all at the same resonant frequency. By inspecting the spectral displacement of a point located at the core of the metabarrier, after $25$ cells ($r_4$), this effect becomes evident. The resonances of the mass-equipped struts interact constructively yielding a wider attenuation zone that extends from $f_r$ to their antiresonance \cite{Colombi2016forests}, as reported by the gray region and red signal in Figure \ref{FIG:MetabarrierConstant}d. 
Furthermore, by comparing this spectrum with the displacement field in a receiver on the edge of the metastructure ($r_3$), black line in Figure \ref{FIG:MetabarrierConstant}d, we observe that the majority of the energy with frequency content within the forbidden range is converted into reflections \cite{Zaccherini2020}. The wavefront impacting on the metabarrier at $7.2$ ms, propagates backward in the plate at \textit{t}=$8.8$ ms (Figure \ref{FIG:MetabarrierConstant}c) after interacting with the resonating members.

Although effectively hampering wave transmission within the bandgap, the constant metawedge exhibits a fixed frequency range of application stemming from the resonant nature of the attenuation zone that, whilst broadband and easily tuneable, entails spatially localized phenomenon. A mitigation device, equipped with smoothly increasing masses, is designed in order to overcome this drawback, as these were proven to affect the frequency content of the prohibited range (see Figure \ref{FIG:DR}c). 
The \textit{graded metabarrier}, with masses $m_1$ to $m_{15}$ ranging from $3m_{ref}$ to $42m_{ref}$, respectively with $t_{mass}$ from $2$ to $5$ mm, is depicted in the right inset of Figure \ref{FIG:System}b and tested with an input Ricker source centered at $3$ kHz. Following the inverse relationship between added mass and resonant frequency, the heaviest masses ($m_{15}$) govern the opening frequency of the total bandgap, while the upper bound depends upon the choice of $m_1$ (see Figure \ref{FIG:MetabarriersGraded}b). The evolution of the bandgap is also shown in Figure \ref{FIG:MetabarriersGraded}b and reveals how the frequencies within the bandgap decrease spatially as masses increase from grade to grade; waves of a certain frequency propagate through the medium until they reach the spatial location where they exist within a bandgap, at this point they are  subsequently trapped and back reflected thus, spatially distillating the spectrum following the Rainbow trapping effect \cite{Tsakmakidis2007, cebrecos2014enhancement,Colombi2016}. As a result, our medium possesses a wider stop band, from $1.14$ kHz to approximately $5.1$ kHz (see gray region in Figure \ref{FIG:MetabarriersGraded}c).  

Figure \ref{FIG:MetabarriersGraded}a illustrates the displacement field filtered over two distinct frequency ranges (see videos in the Supplementary Material). Between $1.8$ and $2.5$ kHz, the wave impinging the metabarrier is almost immediately trapped and scattered by the cells with masses equal to $11m_{ref}$, owing to the matching with their frequency gap ($1.98$-$3.09$ kHz). Similarly, the lower frequency field, corresponding to a pass-band filter between $1.2$ and $1.5$ kHz, travels undisturbed deeper in the metastructure until it eventually interacts with the cells of mass $31m_{ref}$ (bottom inset), the exact location these frequencies occur within the varying bandgap. 

As validation, the spectral displacement of a point at the right end of the metawedge is compared to that of a receiver at an equivalent position in a bare lattice plate in Figure \ref{FIG:MetabarriersGraded}c. The amplitude reduction, stemming from the presence of the metawedge and approximately equal to 87\%, is visibly appreciable, as well as the widening of the bandgap compared to that in Figure \ref{FIG:MetabarrierConstant}d. Notice that, a zero-transmission range exists in both the spectra between $3.8$ and $5.1$ kHz, which can be attributed to the dispersive behavior of the standard octet cells.
While being progressively slowed down within the graded barrier, the impinging energy is trapped by the bending resonant mode of the mass-strut in the octets and consequently amplified. Despite seemingly counterproductive, this dynamic behavior underpins an effective broadband device, where wave attenuation and energy harvesting can coexist. 

\begin{figure*}[t]
    \hspace*{-0.5cm} 
	\centering
		\includegraphics[scale=1]{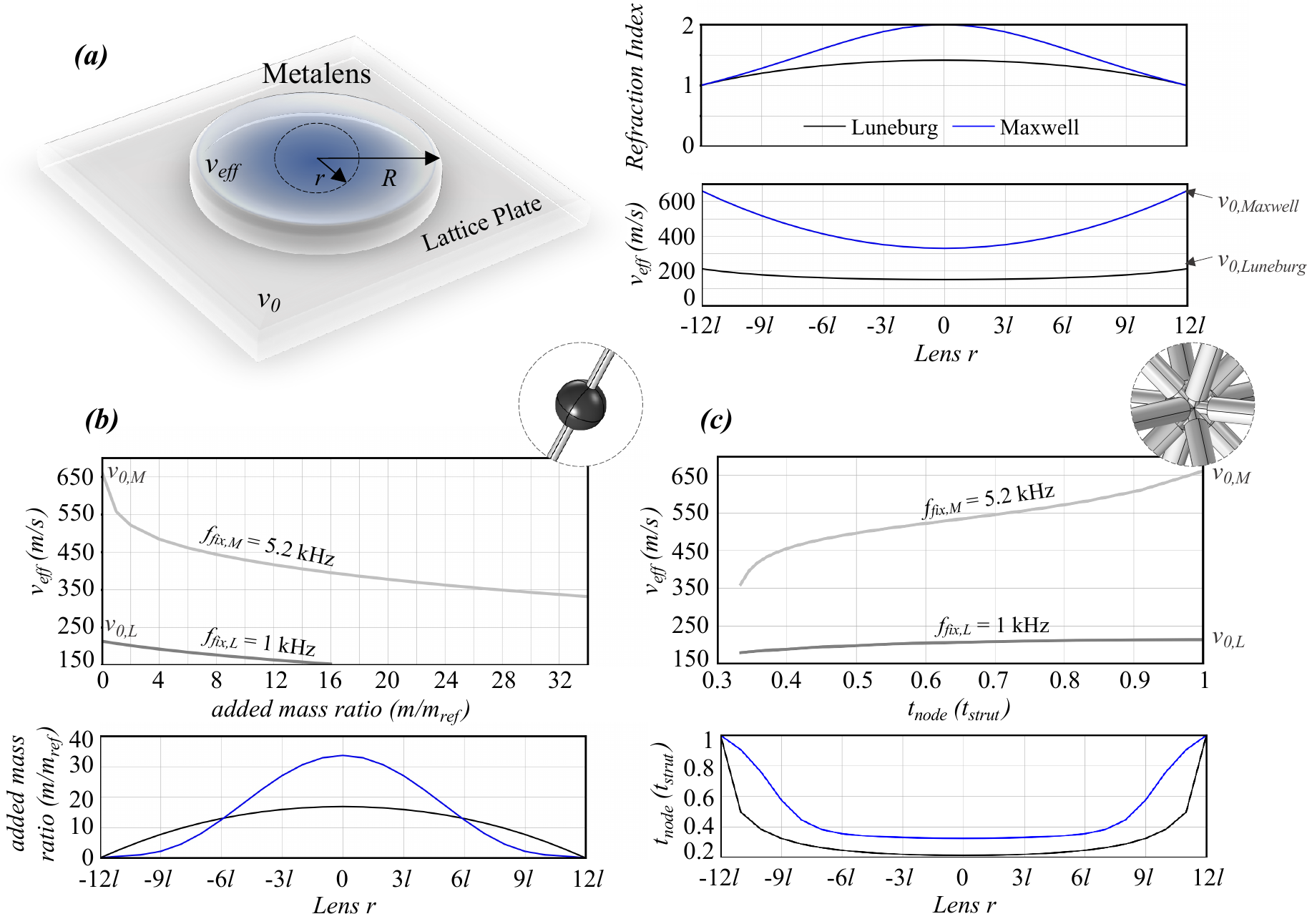}
	\caption{GRIN lenses design. (a) Schematic of the metalenses setup with geometry ($R$=$0.72$m) and phase velocities in the lattice plate, $v_0$, and metadevice, $v_{eff}$, respectively emphasized. The right insert illustrates the index and effective velocity profiles of Luneburg (black) and Maxwell (blue) lenses as a function of the radial coordinate, $r$. $v_{0,Luneburg}$ and $v_{0,Maxwell}$ are the reference velocities of waves in the base lattice plate at $1$ kHz and $5.2$ kHz. At $r$=$R$, $v_{eff}$ = $v_0$. 
	(b) and (c) Design parameters of GRIN lenses comprising the customized octets with auxiliary masses or variable node thickness (see Figures \ref{FIG:SystemDR}c and d). The top insert represents the variation of effective phase velocity with respect to the tuning parameter at the fixed design frequencies of Luneburg and Maxwell lenses. The derived parameters are reported in the bottom inset. }
	\label{FIG:MetalensDesign}
\end{figure*}
\begin{figure*}[t]
	\centering
	 \hspace*{-0.5cm} 
		\includegraphics[scale=1]{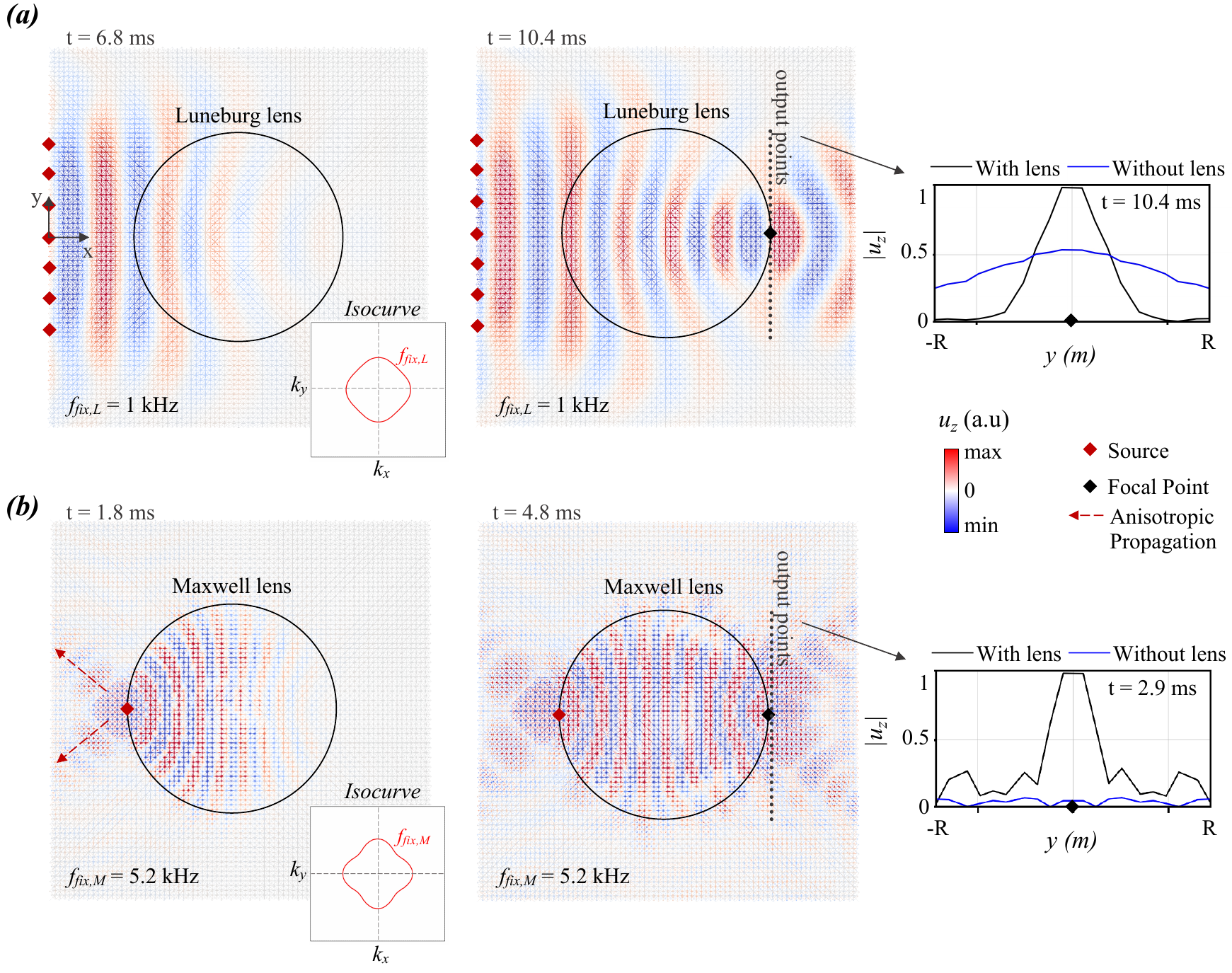}
	\caption{Wave field snapshots in gradient-index lenses consisting of octets with embedded masses. 
	(a) Luneburg lens (black circle) with impinging plane sinusoidal wave of frequency $f_{fix,L}$ at two consecutive instants. (b) Maxwell lens excited by a point source (red diamond) of frequency $f_{fix,M}$ at two instants. 
	In both figures, the isofrequency contours of a standard octet-based plate within the first Brillouin zone at $1$ and $5.2$ kHz are depicted in the bottom right insets. The flexural displacement recorded by a line of receivers is compared to the equivalent field in a bare plate (insets).}
	\label{FIG:Metalenses}
\end{figure*}

\subsubsection*{Octet-based gradient index lenses}
\label{sec32}
Drawing from the dynamic properties of the octet lattice, two of the most popular GRIN lenses are engineered from this architecture to shape ray trajectories and focus wave energy. The design of these devices relies on the dependency of the phase velocity, extracted from the dispersion relations, upon the frequency and the tuning parameters. By combining this relationship with the refractive index profile, the design variables of the customized octet cells are derived at different locations in the lens.

Specifically, the Luneburg and Maxwell lenses are created by gradually varying the supplementary masses embedded in the octets, or the stiffness of their joints, designed to produce the required refractive index profile within a bounded region of radius $R$, which corresponds to $12$ cell layers, each featured by a distinct parameter value (Figure \ref{FIG:System}c). The required theoretical refractive index profile is defined as a function of the radial coordinate, $r$, centred in the middle of the lens, as follows:
\begin{equation}
    n_{Luneburg} = \sqrt{2-(r/R)^2} \quad\text{and}\quad n_{Maxwell} = 2/(1+(r/R)^2)
    \label{eq:RIL}
\end{equation} 
The effective wavespeeds in the metalenses, $v_{eff}$, and in the surrounding lattice plate comprising only standard octets, $v_{0}$, are retrieved from the band structure of the cells, computed via Bloch analysis, with the relationship $v_{eff}$ = $\omega$/$k$. In order to eliminate their frequency dependence, originated by the dispersive nature of the octet, these quantities are evaluated at fixed components, $f_{fix}$. The frequencies are selected sufficiently far from the stop band, so as to prevent the waves from being localized and amplified by the strut resonances. At $1$ kHz and $5.2$ kHz, respectively for the Luneburg ($f_{fix,L}$) and Maxwell ($f_{fix,M}$) lens, the phase velocity is calculated for increasing added masses or node thickness and plotted in the top inset of Figures \ref{FIG:MetalensDesign}b and c. While both wavespeeds feature a decreasing trend for heavier masses, a significant drop can be observed in the high-frequency branch ($f_{fix,M}$), posing as a suitable candidate for the design of a Maxwell lens, which requires an abrupt variation of the refractive index. The opposite behavior is instead noted when varying the thickness of the nodes yielding a speed-up of the waves that propagate from hinge-based ($t_{node}$ = $0.3$ $t_{strut}$) to stiffer lattices ($t_{node}$ = $t_{strut}$). These findings serve as a bridge between the theoretical profiles of GRIN lenses and their actual implementation using reticulated structures, as they relate the octet properties to the refractive index, $n$, by means of the effective velocity.  

After computing the refractive index profiles of the Luneburg and Maxwell lenses with the formula from Equation \ref{eq:RIL}, as shown in Figure \ref{FIG:MetalensDesign}a for increasing radial coordinate, the following expression for the refractive index, 
\begin{equation}
    n(r) = \frac{v_0}{v_{eff}},  
    \label{eq:RefrIndex}
\end{equation} 
is used to retrieve the phase velocity distributions illustrated in the bottom inset of Figure \ref{FIG:MetalensDesign}a. Finally, an inverse design approach, relying on the relations in top Figures \ref{FIG:MetalensDesign}b and c, is adopted to identify the tuning parameters of the customized octets at different $r$, based on the input effective velocity. The designed point masses and node thicknesses are outlined in Figures \ref{FIG:MetalensDesign}b and c for Luneburg and Maxwell lenses. 
Specifically, embedded masses from $3m_{ref}$ to $17m_{ref}$ or node thicknesses ranging from $0.2$ to $1$ $t_{strut}$ are determined according to the Luneburg lens refractive index (from $1$ to $1.41$). The fish-eye lens, instead, originates from a stronger variation of the index profile (from $1$ to $2$), which requires masses from $1$ to $34m_{ref}$ or $t_{node}$ to vary between $0.3$ and $1$ $t_{strut}$. 

These designs, engineered via Bloch theory, are validated by means of time domain numerical simulations of waves traveling in the finite plate of Figure \ref{FIG:System}a, with the Real-ESSI Simulator (see Methods). Both GRIN metalenses are obtained by endowing the octet cells with the masses estimated in Figure \ref{FIG:MetalensDesign}b and are excited by sinusoidal sources tuned at the design frequencies $f_{fix,L}$ and $f_{fix,M}$. As input signal for the Luneburg one, we use a plane wave of $40$ms and $1$kHz. In the Maxwell case we instead use a point-like source of frequency $5.2$ kHz and duration $10$ms. The wave field travels through the lenses with minimal losses and reflections, thanks to the gradual variation of properties from the plain octet lattice at $r$ = $R$ to the center of the device that minimize the impedance mismatch (see videos in the Supplementary Material). In Figure \ref{FIG:Metalenses}a, a plane wave impinging on the left side of the Luneburg lens at $6.8$ ms is progressively bent by the smooth refractive index transition and its trajectory is steered toward a diametrically opposite focal point placed on its surface ($t$=$10.4$ms).
An array of receivers, positioned along this edge, outlines the significant energy localization generated by the lens. A sharp spike emerges in the out-of-plane displacement $u_z$ and its peak value is almost twice as large as its counterpart on the bare plate. 

Similarly, two distinct time instants in Figure \ref{FIG:Metalenses}b validate the Maxwell lens design, where the wave originating from a point source located at the leftmost side of the lens is guided toward an antipodal focus thus, creating the typical fish-eye effect. Notice that, a non-circular wavefront is produced by the fields of the source and focal point in the surrounding plate, which instead propagate along preferential directions (approximately at \ang{45}) due to the anisotropic character of the standard octet lattice. This behavior is further corroborated by the isofrequency contours of the standard octet lattice at $f_{fix,M}$, reported in the bottom inset of Figure \ref{FIG:Metalenses}b. A $70$x$70$ plate, complemented with absorbing boundary conditions and a broadband point source placed at its center, is leveraged for the numerical derivation of the isocurves. The resulting wave field is spatio-temporally Fourier transformed and the snapshot corresponding to $5.2$ kHz is then extracted. The circular isofrequency contours, typical of an isotropic medium, distort here to a rhombic shape emphasizing the anisotropic nature of the lattice. The wave vectors impinging any of these contours are transmitted along the diagonals of the first Brillouin zone, causing the wave field in the real space to assume the cross-like form that emerges in Figure \ref{FIG:Metalenses}b. A similar anisotropic trait can be appreciated at lower frequencies ($f_{fix,L}$ =$1$ kHz) upon inspection of the corresponding isofrequecy contour (bottom inset of Figure \ref{FIG:Metalenses}a). 
In addition, the anisotropy explains the near-zero vertical displacement in the receivers of the bare plate, blue curve in the inset of Figure \ref{FIG:Metalenses}b, which juxtaposes with the peak of the highly focused signal from the lens (black line). 
Equivalent results are obtained when the lens design is based on the variation of the node thickness. The only difference lies in the shape of the metadevice, squared instead of circular, to avoid discontinuities in the boundary conditions of the struts. 

\section*{Discussion}
We have demonstrated the potential of octet-based lattice plates in controlling the trajectories of asymmetric Lamb-like waves, be it via attenuation or wave guiding. 
Starting from the cellular architecture of this structure, we prove the standard octet unit is capable of supporting a Fano-type bandgap ($3.85$-$5.08$ kHz) induced by the coalescence of local bending resonances and Bragg frequency, opening the avenue for an easy-to-tune, single-phase material in elastic applications. The design of this woven structure is simplified by viewing the struts as Euler-Bernoulli beams clamped at both their ends and leveraging the analytical expression of the fundamental flexural mode as a prediction paradigm for the bandgap opening frequency; furthermore, since the octet contains struts of length and direction identical to that of the primitive lattice vectors, it creates a medium in which the clamped-clamped fundamental resonance exists as a standing wave between nodes, hence it is a medium in which local bending resonances and Bragg frequencies coincide, resulting in destructive interferences and a broadband resonant based bandgap.

The two tuning parameters extrapolated from this analogy, midpoint added masses and joint thickness, are proven to control the frequency content of the attenuation zone as well as the slope of the low and high-frequency bands exhibiting out-of-plane polarization. The opening frequency of the bandgap is inversely correlated with the masses and directly correlated with the node thickness; heavier masses drive the forbidden range to lower frequencies and reduce the phase velocity of waves in the lattice, conversely, thicker nodes shift the content of the attenuation zone to higher values and privilege the transmission of faster waves.

Drawing from these intrinsic dynamic behaviors, four devices for controlling ray trajectories are successfully engineered. The devices operate over a range of scales,  the metabarriers leverage the bandgap and hence operate within a Bragg-scattering regime; on the other hand, the Luneburg lens is based on spatial variations of the low-frequency $A_{0}$ mode and hence the underlying octet cells behave in a subwavelength manner. 

The constant metabarrier confirms the existence of a hybridization bandgap ($1.8$-$2.8$ kHz) stemming from the interaction between the propagating flexural field and octet units equipped with auxiliary masses. The wave impinging the resonant members is backscattered and prevented from being transmitted within the metastructure. Whilst the bandwidth of the constant metabarrier is broad, the bandgap and frequency range is spatially localized; we introduce a graded design capable of, not only tripling the mitigation width ($1.14$-$5.1$ kHz), but also spatially filtering the spectrum via a rainbow trapping mechanism, amplifying a large spectrum of frequencies within its resonating struts. The generating mechanism of the stop band, initially exploited by the lattice metabarriers to match the frequency content of incoming fields and inhibit them, serves to also tune the phase velocity within a bounded region and enable wave steering. Two GRIN lenses, the Luneburg and Maxwell fish-eye, are successfully implemented by gradually adjusting the properties of the customized octets according to the theoretical index spatial distribution, thus paving the way for novel applications of this face-centered cubic topology in wave focusing.

Although the dynamic assessment of three-dimensional woven lattices is still at a primordial stage compared to the existing knowledge on their static mechanical properties, the outcomes of our work have shed light on the hidden potential of the octet and tuning parameters. The ease of scaling to different wavelengths that characterizes this structure renders it even more attractive for applications in a variety of domains, including vibration isolation or conversion of focused energy. Indeed, the vibration energy trapped and focused, respectively by the struts and metalenses, could be harvested and converted into other forms of energy, e.g., electrical, by attaching piezoeletric components to the lattice. 
Future steps could, for instance, address the complicated matter of stress concentration due to dynamic loading, as well as the experimental validation of the proposed structures or the extension to different lattice unit cell based both on struts and sheets such as foams.

\section*{Methods}

\textbf{Bloch Analysis.} The dispersion relation, reported in Figure \ref{FIG:DR}a (red dotted line), is solved using the finite element method, performed by the software COMSOL Multiphysics$^\circledR$, by sweeping through the required wavevector $\textbf{k}$ determining the permitted frequencies $\omega = \omega( \textbf{k})$ as an eigenvalue and the displacement field as an eigenvector. 
Our choice of fundamental cell (Figure \ref{FIG:SystemDR}a) is a conventional cubic-like unit cell over a primitive one \cite{Ashcroft1976}, which allows us to align the base vectors with the propagation direction of interest. This makes the comparison between the band structure stemming from the FFT analysis performed on the time-transient finite lattice section (Figure \ref{FIG:SystemDR}b) and the eigenmodal analysis straightforward as we only know the displacement fields along the beams and not everywhere in the cell.
With this choice of unit cell the ends of the struts either correspond to a node point or lie on the edges of the cell, we apply either continuity conditions or quasi-periodic Bloch conditions at the strut ends as appropriate; hence the displacement field takes the following well known form:
\begin{equation}
    \textbf{u} (\textbf{x}) = \textbf{U}( \textbf{x}) e^{ 2 \pi i( \textbf{k} \cdot \textbf{x}- \omega t)}, 
\end{equation}
where $\textbf{k} = \kappa_{x} \textbf{e}_{x} + \kappa_{y} \textbf{e}_{y} + 0 \textbf{e}_{z}$  denotes the in-plane Bloch wavevector, $\omega$ the frequency assuming harmonic time-dependence  and $\textbf{U}(\textbf{x})$ is a periodic function satisfying:
\begin{equation}
    \textbf{U}(\textbf{x} + n \textbf{a}_{1} + m \textbf{a}_{2}) =  \textbf{U}(\textbf{x}),
\end{equation}
for any integers $n$ and $m$. We restrict our attention to waves propagating along the $x$-axis, hence consider only $\textbf{k}$ along the $\Gamma$-$X$ portion of the irreducible Brillouin zone ($0$ - $\frac{1}{2l}$ 1/m).
The struts are modeled as EB beams with 6 degrees of freedom in each node (inset in Figure \ref{FIG:SystemDR}a), reducing the high computational cost of the simulations that would arise when using solid elements. The EB theory was selected over Timoshenko theory upon fulfilling the condition $EI/0.88AGl_{strut}^2<<1$, with $A$ and $I$ area and moment of inertia and $G$ shear modulus, as the slenderness of the beams makes shear deformations negligible. \\

\textbf{Time-transient Simulations.}
The black portion in the band structure (Figure \ref{FIG:DR}a) stems from time-transient simulations of waves propagating in the finite plate shown in Figure \ref{FIG:SystemDR}b, by means of the finite element software Real-ESSI Simulator \cite{RealEssi}. 
The system is excited by a modulated plane sweep source ranging from $0.1$ to $9$ kHz with components in all three directions of the space. The asymmetric Lamb waves propagating in the plate are recorded by an array of points lying perpendicularly to the source along the metabarrier center (blue dotted line in Figure \ref{FIG:SystemDR}b) and the resulting wave field $u_z$ is spatio-temporally Fourier transformed to derive the correspondent band structure.

The metadevices (Figure \ref{FIG:System}) are tested via $3$D time-transient analyses conducted with the parallel version of the Real-ESSI Simulator \cite{RealEssi} in order to substantially reduce the computational time. The Newmark integration method is employed with $\gamma$ equal to 0.505. The parameter $\gamma$ controls the numerical or algorithmic damping, that is inserted into the model to damp out any non-realistic high frequencies stemming from model discretization ($\gamma$ equal to 0.5 would correspond to zero numerical damping). It should be noted, here, that this type of damping cannot be used to represent any physical damping mechanism (e.g. viscous damping, hysteretic damping), while improper use could lead to unrealistic attenuation of the real response of the model \cite{Constantinos2020} (i.e. values greater than 0.5 should be used with caution). External software are employed for both pre- and post-processing. Specifically, the geometry and mesh (Figure \ref{FIG:System}a) are generated with the 3D finite element generator Gmsh and a translator, called gmESSI, is invoked to convert the mesh file from Gmsh to a format readable by Real-ESSI. The mesh of the struts in the lattice consists of three-dimensional EB beam elements. The customized cells in Figures \ref{FIG:SystemDR}c and d can be easily implemented either by adding point masses to the center of the struts or locally modifying their cross-section in the node area, condensing the degree of complexity to the design stage with Gmsh. Although the Real-ESSI Simulator output results can be visualized with the open-source application Paraview after loading the PVESSIReader plugin \cite{RealEssi}, for this study an in-house code, written in Python, was employed to postprocess and later visualize the wave field in Paraview.

The in-plane size of the full frame plate is specified by the number of cells, $N_{cells,x}$ and $N_{cells,y}$ in Figure \ref{FIG:System}a, multiplied by the cell width, $l$. This domain is tested under vertically polarised excitations with unit amplitude that entail flexural Lamb waves propagating along $x$. Depending on the assessed metastructure, the input may be a line, for metabarriers and Luneburg lens, or punctual source, for Maxwell lens, and its frequency content span from that of a pulse-like Ricker excitation to a monochromatic source (inset of Figure \ref{FIG:System}a). Absorbing layers using quadratically increasing Rayleigh damping (ALID) \cite{Rajagopal2012}, defined along the five outer cell arrays of the plate, are implemented to mimic an unbounded domain thus, minimizing undesirable reflections from the boundaries. 
A total of twenty absorbing layers on every edge is considered, each of thickness $l/4$, and a mass proportional damping with maximum coefficient $c_{max}$ equal to $2\cdot10^4$ is assumed. The inner portion of the plate is, instead, left damping-free to avoid compromising the resonance of the struts and given that the selected material, versatile plastic, is characterized by a low quality factor (Q=$50$).

\bibliography{Bibliography}

\section*{Acknowledgements}
GA and AC were supported by the H2020 FETOpen project BOHEME under grant agreement No. 863179 and by the Ambizione Fellowship PZ00P2-174009.
CK was supported by  the H2020 "INSPIRE" EU program under grant agreement No 813424. RW thanks the UK EPSRC for their support through grant EP/L016230/1.

\end{document}